\documentclass[aps,prd,twocolumn,showpacs,nofootinbib]{revtex4-1}

\usepackage{graphicx}
\usepackage[colorlinks=true, linkcolor=blue, citecolor=blue, urlcolor=blue]{hyperref}

\begin{document}

\title{Hadronic production of the doubly charmed baryon via the proton-nucleus and the nucleus-nucleus collisions at the RHIC and LHC }

\author{Gu Chen$^1$}
\email{speecgu@gzhu.edu.cn}
\author{Chao-Hsi Chang$^{2,3}$}
\email{zhangzx@itp.ac.cn}
\author{Xing-Gang Wu$^4$}
\email{wuxg@cqu.edu.cn}

\address{$^1$School of Physics $\&$ Electronic Engineering, Guangzhou University, Guangzhou 510006, People's Republic of China\\
$^2$Institute of Theoretical Physics, Chinese Academy of Sciences, P.O.Box 2735, Beijing 100080, People's Republic of China\\
$^3$ School of Physical Sciences, University of Chinese Academy of Sciences, Beijing 100049, China\\
$^4$Department of Physics, Chongqing University, Chongqing 401331, People's Republic of China  }

\date{\today}

\begin{abstract}

We present a detailed discussion on the doubly charmed baryon $\Xi_{cc}$ production at the RHIC and LHC via the proton-nucleus ($p$-N) and nucleus-nucleus (N-N) collision modes. The extrinsic charm mechanism via the subprocesses $g+c\to (cc)[n]+\bar{c}$ and $c+c\to (cc)[n]+g$ together with the gluon-gluon fusion mechanism via the subprocess $g+g\to(cc)[n]+\bar{c}+\bar{c}$ have been taken into consideration, where the intermediate diquark is in $[n]=[^1S_0]_{\bf 6}$-state or $[^3S_1]_{\bar{\bf 3}}$-state, respectively. Total and differential cross sections have been discussed under various collision energies. To compare with the $\Xi_{cc}$ production via proton-proton collision mode at the LHC, we observe that sizable $\Xi_{cc}$ events can also be generated via $p$-N and N-N collision modes at the RHIC and LHC. For examples, about $8.1\times10^7$ and $6.7\times10^7$ $\Xi_{cc}$ events can be accumulated in $p$-Pb and Pb-Pb collision modes at the LHC within one operation year. \\


\end{abstract}

\maketitle

\section{Introduction}

Within the quark model, the doubly heavy baryon is regarded as a three quark state with two heavy quarks ($c$ or $b$) and a light quark $q$ ($q=u, d, s$)~\cite{GellMann:1964nj, Zweig:1981pd, Ebert:1996ec, Gerasyuta:1999pc, Itoh:2000um}. The doubly heavy baryons are important for the understanding of Quantum Chromodynamics (QCD) theory. For convenience, throughout the paper, we adopt $\Xi_{QQ'}$ as short notation for the baryon $\Xi_{QQ'q}$, where $Q$ and $Q'$ stand for the heavy $c$ or $b$ quark, respectively.

In year 2000, the SELEX collaboration~\cite{Mattson:2002vu, Ocherashvili:2004hi} reported the observation of $\Xi_{cc}^+$ via its decay channels $\Xi_{cc}^+ \to \Lambda_c^+K^-\pi^+$ and $\Xi_{cc}^+ \to pD^+K^-$. Later one, more experimental measurements have been carried out to confirm this observation by the FOCUS~\cite{Ratti:2003ez}, the BABAR~\cite{Aubert:2006qw}, the Belle~\cite{Chistov:2006zj,Kato:2013ynr}, and the LHCb collaboration~\cite{Aaij:2013voa}. However, all of those experiments were fail to reproduce the SELEX observation. In year 2017 the LHCb collaboration released their first observation of $\Xi_{cc}^{++}$~\cite{Aaij:2017ueg} via its weak decay channel $\Xi_{cc}^{++} \to \Lambda_c^+K^-\pi^+\pi^+$. This stimulates many new works, either experimentally or theoretically, for the doubly heavy baryons. The LHCb observation is based on the simulation by using a dedicated generator GENXICC~\cite{Chang:2007pp, Chang:2009va, Wang:2012vj}, which is designed to simulate the doubly heavy baryon production via proton-proton ($pp$) collision at the Large Hadron Collider (LHC).

Theoretically, the production of doubly heavy baryons at different types of high-energy colliders, such as the $e^+e^-$, the electron-proton ($ep$), and the $pp$ (or $p\bar{p}$) colliders, has been studied in detail in many works, cf. Refs.\cite{Falk:1993gb, Kiselev:1994pu, Baranov:1995rc, Berezhnoy:1998aa, Gunter:2001qy, Braguta:2002qu, Braaten:2003vy, Ma:2003zk, Chang:2006eu, Chang:2006xp, Li:2007vy, Yang:2007ep, Zhang:2011hi, Jiang:2012jt, Jiang:2013ej, Martynenko:2013eoa, Chen:2014frw, Yang:2014tca, Yang:2014ita, Martynenko:2014ola, Leibovich:2014jda, Chen:2014hqa, Zheng:2015ixa, Brodsky:2017ntu, Yao:2018zze}. At the hadronic colliders, those works culminated in the generator GENXICC. This generator not only contains gluon-gluon fusion and the extrinsic charm  mechanisms via the $g+g\to\Xi_{cc}+X$, $g+c\to\Xi_{cc}+X$ and $c+c\to\Xi_{cc}+X$ subprocesses, but also contains various contributions from high Fock states in the baryons \footnote{The general-mass variable-flavor-number scheme~\cite{Aivazis:1993kh, Aivazis:1993pi, Olness:1997yc, Amundson:2000vg} has been adopted to deal with the double coupling problem among different channels and the QCD factorization is done within the framework of nonrelativistic QCD (NRQCD)~\cite{Bodwin:1994jh}.}.

Similar to the case of $B_c$-meson production~\cite{Aarts:2016hap, Chang:2015hqa, Chen:2018obq}, it is interesting to show that whether sizable number of $\Xi_{cc}^{++}$ events can also be produced at the heavy ion colliders, such as the STAR experiment at the Relativistic Heavy Ion Collider (RHIC) and a large ion collider experiment (ALICE) at the LHC. In the heavy ion collisions, the $\Xi_{cc}^{++}$ baryons can be produced via the proton-nucleus ($p$-N) and the nucleus-nucleus (N-N) collision modes, respectively. The production of doubly heavy baryon via $p$-N or N-N collision should also provide an alternative candidate of studying on the quark-gluon plasma (QGP), similar to the case of doubly heavy meson production via heavy ion collision~\cite{Chen:2018obq}. Through high-energy $p$-N and N-N collisions, the generated heavy quarks will either combine and evolve into doubly heavy baryon before QGP formation or hadronize into colorless baryons via the transition from the QGP phase to the hadronic phase. This makes the production of $\Xi_{cc}^{++}$ baryons via the $p$-N and N-N modes quite different from the usual $pp$-collision mode. Especially, the nuclear effects, e.g., the shadowing effect and the modifications of nuclear parton density functions (nPDFs) etc., shall play a significant role. Consequently, by measuring the properties of the doubly heavy baryon, one may achieve important information about QGP and nuclear properties.

The remaining parts of the paper are organized as follows. In Sec.II, we present the calculation technology. In Sec.III, we present the numerical results for the $\Xi_{cc}^{++}$ production via $p$-N or N-N collision model, and the results for the $pp$ collision mode are also presented for comparison. Sec.IV is reserved for a summary.

\section{Explanation of the calculation technology}

The production of $\Xi_{cc}$ baryon can be factorized into two steps: The first step is to produce two $c\bar{c}$ pairs. This step is pQCD calculable, since the intermediate gluon should be hard enough to generate a heavy $c\bar{c}$ pair. The second step is to make the heavy $c$-quarks into a bounding $(cc)$-diquark in the spin-triplet ($[^3S_1]$) or spin-singlet ($[^1S_0]$), and the $\mathbf{\bar{3}}$ (or $\mathbf{6}$) color state, accordingly. More explicitly, the intermediate diquark in $\Xi_{cc}$ has two spin-and-color configurations, $[^3S_1]_{\bf\bar{3}}$ and $[^1S_0]_{\bf 6}$. It will then be hadronized into a $\Xi_{cc}$ baryon via fragmentation, whose probability is characterized by the non-perturbative matrix element: 1) We adopt the usual assumption that the diquark shall evolve into the baryon with $100\%$ probability~\footnote{By using a concrete fragmentation function, as suggested by Ref.\cite{Peterson:1982ak}, to deal with the production shall give quite small difference to this simple assumption~\cite{Chen:2014frw}.}; 2) The intermediate $(c\bar{c})$-diquark shall grab a light quark with possible soft gluons from the hadron to form the final colorless baryon with a relative possibility for different light quarks as $u :d :s \simeq 1:1:0.3$~\cite{Sjostrand:2006za}. If the diquark $(cc)[^3S_1]_{\bar{3}}$ is produced, then it will fragment into $\Xi_{cc}^{++}$ with $43\%$ probability, $\Xi_{cc}^{+}$ with $43\%$ probability, and $\Omega_{cc}^{+}$ with $14\%$ probability.

Consequently, the production of $\Xi_{cc}$ baryon can be expressed as~\cite{Bodwin:1994jh},
\begin{widetext}
\begin{eqnarray}\label{factexp}
d\sigma({\rm A}{\rm B} \to {\Xi_{cc}}+X) &=&  \sum_{i,j=g,c}\sum_{n} \int dx_1 dx_2 (N_{\rm A} f^{h/{\rm A}}_{i}(x_1, \mu_{f}))  (N_{\rm B} f^{h/{\rm B}}_{j}(x_2, \mu_{f})) d\hat\sigma_{ij \to (cc)[n]+ X} \langle{\cal O}^{\cal H}[n] \rangle,
\end{eqnarray}
\end{widetext}
where the symbol $[n]$ stands for the spin-and-color state of the $(cc)$-diquark, and the symbols $\rm A$ and $\rm B$ stand for $p$ or N for the incident hadron to be proton or nucleus, respectively. The functions $(N_{\rm A} f_{i, j}^{h/\rm A})$ and $(N_{\rm B} f_{i, j}^{h/\rm B})$ with $(i,j=g,c)$ are effective nucleus parton distribution functions (PDFs) for the nucleus $\rm A$ or $\rm B$ accordingly, which stands for the parton density of bound-nucleon $h$ in nucleus A and carries the fraction $x_n\,(n=1,2)$ of the hadron momentum at the factorization scale $\mu_f$. $h$ stands for the nucleon, proton or neutron, respectively. Here $N_{\rm A}$ or $N_{\rm B}$ is the nuclear number in the incident nucleus. For examples, $N_{\rm Au}=197$ for the gold nucleus ($^{197}_{79}\rm Au$), and $N_{\rm Pb}=208$ for the lead nucleus ($^{208}_{82}\rm Pb$). Many PDF models have been suggested to study the heavy-ion collisions, such as the Heavy-Ion Jet INteraction Generator (HIJING) model~\cite{Wang:1991hta}, A Multiphase Transport (AMPT) model~\cite{Lin:2004en}, the Monte Carlo Glauber Model~\cite{Alver:2008aq, Broniowski:2007nz, Rybczynski:2013yba, Loizides:2017ack}, and etc.. Here, following the same idea of CTEQ group~\cite{Kovarik:2015cma}, we adopt the PDF of bound-nucleon in nucleus as the heavy ion PDF.

For the $pp$ collision mode, we have $N_{\rm A}=N_{\rm B}=1$, and the PDFs are reduced to the usual PDFs inside the free proton. For the case of $N_{\rm A}\neq 1$ and $N_{\rm B}\neq1$, we need to consider the collision geometry and the spatial dependence of the shadowing parameterization effect~\cite{Aubert:1987da, Gavin:1990gm} to the nucleus PDF. More explicitly, for the case of $p$-N and N-N collisions, we need to know the nuclear gluon/charm distribution functions. For the nuclear A, its overall gluon/charm distribution function can be expressed as
\begin{eqnarray} \label{nucleus}
f^{\rm A}_{g(c)}(x_n, \mu_f) = \int d^2 \vec{r} dz  \,f^{\rm A}_{g(c)}(x_n, \mu_{f}, \vec{r}, z),
\end{eqnarray}
where $\vec{r}$ and $z$ are transverse and longitudinal location of the parton in the coordinate space. Considering the nuclear effects and collision geometry, the nuclear densities $f^{\rm A}_{g(c)}(x_n, \mu_f, \vec{r}, z)$ can be factorized as the product of the nucleon density in the nucleus $\rho_{\rm A}(s)$, the free-nucleon density $f^{h}_{g(c)}(x_n, \mu_f)$, and the shadowing ratio $S^{g(c)}_{\rm P, S}(N_{\rm A}, x_n, \mu_f, \vec{r}, z)$, i.e.
\begin{eqnarray}
&& f^{\rm A}_{g(c)}(x_n, \mu_f, \vec{r}, z)  \nonumber\\
&=& \rho_{\rm A}(s) S^{g(c)}_{\rm P, S}(N_{\rm A}, x_n, \mu_f, \vec{r}, z) f^{h}_{g(c)}(x_n, \mu_f), \label{npdf}
\end{eqnarray}
where $s=\sqrt{r^2+z^2}$, $\rho_{\rm A}(s)$ is assumed to be Woods-Saxon distribution~\cite{DeJager:1974liz}, which satisfies the normalization condition
\begin{eqnarray}
 \int \rho_{\rm A}(s) d^2\vec{r} dz  &=& N_{\rm A}.
\end{eqnarray}
Then, we obtain
\begin{eqnarray}
 && \int d^2 \vec{r} dz \rho_{\rm A}(s) S^{g(c)}_{\rm P, S}(N_{\rm A}, x_n, \mu_f, \vec{r}, z) \nonumber \\
 &=& N_{\rm A}\,S_{\rm P}^{g(c)}(N_{\rm A}, x_n, \mu_f).
\end{eqnarray}
Consequently, we have~\cite{Vogt:2010aa, Brambilla:2004wf}
\begin{eqnarray}
f^{\rm A}_{g(c)}(x_n, \mu_f) &=& N_{\rm A} \,S_{\rm P}^{g(c)}(N_{\rm A}, x_n, \mu_f) f^h_{g(c)}(x_n, \mu_f)   \nonumber\\
                                              &=& N_{\rm A}\,f_{g(c)}^{h/\rm A}(x_n, \mu_f).   \label{fnaln}
\end{eqnarray}
Here $f_{g(c)}^{h/\rm A}(x_n, \mu_f)$ is the wanted effective bound PDF for nucleon, which describes the gluon/charm density of the bound-nucleon $h$ in nucleus A. We adopt the nCTEQ15 version to calculate $f_{g(c)}^{h/\rm A}(x_n, \mu_f)$, which is fixed via a global fit by using the experimental data on nuclei all the way up to $^{208}\rm Pb$~\cite{Kovarik:2015cma}.

Furthermore, the $\Xi_{cc}$ baryon can be expanded as a series of Fock states over the relative velocity ($v$) of the constituent heavy quarks in the baryon rest frame,
\begin{displaymath}
\vert\Xi_{cc} \rangle = c_1 \vert (cc) q\rangle +c_2 \vert (cc) qg \rangle +c_3 \vert (cc) q gg \rangle +\cdots,
\end{displaymath}
where the expansion coefficients $c_i(i=1,2,\cdots)$ are functions of $v$. $\langle{\cal O}^{\cal H}[n] \rangle$ is the long-distance matrix element, which is proportional to the inclusive transition probability of the perturbative diquark state $(cc)[n]$ into the heavy baryon $\Xi_{cc}$. For convenience, we adopt the assumption that the transition probability for the color anti-triplet or color sextuplet of the $(cc)$-diquark are the same~\cite{Ma:2003zk}. The non-perturbative long-distance matrix element can be related to the Schr$\ddot{\rm o}$dinger wavefunction at the origin as~\cite{Chang:2006eu}, $\langle{\cal O}^{\cal H}[n] \rangle \simeq |\psi_{cc}(0)|^2$. $d\hat\sigma_{ij \to (cc)[n]+ X}$ is the differential cross-section of the hard subprocess, which are different for different channels and shall be dealt with by using the generator GENXICC.

\section{Numerical results and discussions}\label{results}

To do the numerical calculation, we take $|\Psi_{cc}(0)|^2 =0.039$ GeV$^3$~\cite{Baranov:1995rc}, $M_{\Xi_{cc}}=3.50$ GeV with $m_c=M_{\Xi_{cc}}/2$, and the nCTEQ15~\cite{Kovarik:2015cma} as the nucleon PDF. The renormalization scale and the factorization scale are set to be the transverse mass of $\Xi_{cc}$, $M_t=\sqrt{M^2_{\Xi_{QQ'}}+p_t^2}$. For the collision energies, we adopt~\cite{Chen:2018obq, PDG}: $\sqrt{S_{p\rm Pb}}= 8.16\;\rm TeV$ and $\sqrt{S_{\rm PbPb}}= 5.02\;\rm TeV$ at the LHC, and $\sqrt{S_{p\rm Au}}= 0.2\;\rm TeV$ and $\sqrt{S_{\rm AuAu}}= 0.2\;\rm TeV$ at the RHIC.

\subsection{Basic results for $\Xi_{cc}$ production}

\begin{table*}[htb]
\centering
\begin{tabular}{|c|c|c|c|c|c|}
\hline
- & \multicolumn{2}{|c|}{RHIC} & \multicolumn{3}{|c|}{LHC} \\
\hline
$\sqrt{S_{\rm NN}}$ (TeV)  & $p$-Au (0.2) & Au-Au (0.2) & $pp$ (13)  & $p$-Pb (8.16) & Pb-Pb (5.02)  \\
\hline
$\sigma(gg\to (cc)_{\bf 6}[^1S_0])$ & $1.05\times10^{-1}$ & $2.30\times10^{1}$ & $7.96\times10^{-2}$  & 9.20 & $1.04\times10^{3}$ \\
\hline
$\sigma(gg\to (cc)_{\bar{\bf 3}}[^3S_1])$ & $5.79\times10^{-1}$ & $1.27\times10^{2}$ & $4.19\times10^{-1}$ & $4.79\times10^{1}$ & $5.38\times10^{3}$  \\
\hline
$\sigma(gc\to (cc)_{\bf 6}[^1S_0])$ & $2.83\times10^{-1}$ & $6.26\times10^{1}$ & $9.40\times10^{-2}$  & $1.17\times10^{1}$ & $1.19\times10^{3}$  \\
\hline
$\sigma(gc\to (cc)_{\bar{\bf 3}}[^3S_1])$ & 2.58 & $5.73\times10^{2}$ & $8.64\times10^{-1}$ & $9.31\times10^{1}$ & $1.09\times10^{4}$  \\
\hline
$\sigma(cc\to (cc)_{\bf 6}[^1S_0])$ & $4.06\times10^{-5}$ & $9.00\times10^{-3}$ & $8.80\times10^{-5}$ & $1.05\times10^{-2}$ & 1.25  \\
\hline
$\sigma(cc\to (cc)_{\bar{\bf 3}}[^3S_1])$ & $1.01\times10^{-3}$ & $2.23\times10^{-1}$ & $2.33\times10^{-3}$ & $2.94\times10^{-1}$ & $3.33\times10^{1}$  \\
\hline
\end{tabular}
\caption{Total cross sections (in unit $\mu$b) for $\Xi_{cc}$ in $pp$, $p$-N, and N-N collision modes at the RHIC and the LHC.} \label{totcro}
\end{table*}

In Table~\ref{totcro}, we present the total cross sections for the production of $\Xi_{cc}$ via the $p$-N and N-N collision modes at the RHIC and the LHC. By summing up contributions from $g+g$, $g+c$ and $c+c$ channels and different spin-and-color configurations of the intermediate $(cc)$-diquark together, we obtain
\begin{eqnarray}
\left.\sigma^{\rm tot}_{p\rm Au}(\Xi_{cc})\right |_{\rm RHIC} &=& 3.55\;{\rm \mu b},  \\
\left.\sigma^{\rm tot}_{\rm AuAu}(\Xi_{cc}) \right|_{\rm RHIC} &=& 7.85\times10^{2}\;{\rm \mu b}, \\
\left.\sigma^{\rm tot}_{pp}(\Xi_{cc})\right |_{\rm LHC} &=& 1.46\;{\rm \mu b},  \\
\left.\sigma^{\rm tot}_{p\rm Pb}(\Xi_{cc})\right |_{\rm LHC} &=& 1.62\times10^{2}\;{\rm \mu b},  \\
\left.\sigma^{\rm tot}_{\rm PbPb}(\Xi_{cc})\right |_{\rm LHC} &=& 1.85\times10^4\;{\rm \mu b}.
\end{eqnarray}
To compare with $\Xi_{cc}$ production via the $pp$ collision, we observe that the total cross sections of $\Xi_{cc}$ are enhanced by about $2 \sim 4$ orders of magnitude in $p$-N and N-N collision modes at the RHIC and  the LHC.

At the RHIC, the designed luminosities are $4.5\times 10^{29}\,\rm cm^{-2}s^{-1}$ and $8.0\times 10^{27}\,\rm cm^{-2}s^{-1}$ for the $p$-Au and the Au-Au collisions \footnote{In predicting the event numbers, at the RHIC, one operation year means $10^7$s for $p$-Au and Au-Au collisions; At the LHC, one operation year means $10^7$s for $pp$-collision and $10^6$s for $p$-Pb and Pb-Pb collisions~\cite{Carminati:2004fp, Alessandro:2006yt}.}. Thus, at the RHIC, we shall have $1.6 \times 10^7$ $\Xi_{cc}$ events to be generated for the $p$-Au collision in one operation year, $6.3 \times 10^7$ $\Xi_{cc}$ events for the Au-Au collision. At the LHC, the designed luminosities are $5.0\times 10^{33}\,\rm cm^{-2}s^{-1}$, $5.0\times 10^{29}\,\rm cm^{-2}s^{-1}$ and $3.6\times 10^{27}\,\rm cm^{-2}s^{-1}$ for the $pp$, $p$-Pb and the Pb-Pb collisions. Thus, at the LHC, we shall have $7.3 \times 10^{10}$ $\Xi_{cc}$ events to be generated in $pp$ collision in one operation year, $8.1 \times 10^7$ $\Xi_{cc}$ events to be generated in $p$-Pb collision, $6.7 \times 10^7$ $\Xi_{cc}$ events for the Pb-Pb collision. Total cross-sections at the RHIC and LHC are in the following sequential order, $\sigma_{pp}<\sigma_{p{\rm A}}<\sigma_{\rm AA}$; due to the shadowing effect, the relative ratio among those cross-sections is smaller than the naive ratio, $1:N_{\rm A}:N^2_{\rm A}$, with A=Au for RHIC and A=Pb for LHC. At the LHC, the $\Xi_{cc}$ events generated via $p$-Pb and Pb-Pb collisions are about three-orders lower than those via $pp$ collision, which are still sizable; thus the STAR and ALICE experiments at the RHIC and LHC can also be potential platform for studying the properties of $\Xi_{cc}$.

For the $p$-Pb ($p$-Au) collision at the LHC (RHIC), Table~\ref{totcro} shows that the contribution from $(cc)_{\bf \bar{3}}[^3S_1]$ is larger than that of $(cc)_{\bf 6}[^1S_0]$ by about five (six) times for $g+g$-mechanism, eight (nine) times for $g+c$-mechanism, and twenty-eight (twenty-five) times for $c+c$-mechanism. The conditions for the N-N collision are similar. Even though the contribution from $(cc)_{\bf 6}[^1S_0]$ is small, one should take it into consideration for an accurate prediction. By summing up the contributions from $(cc)_{\bf 6}[^1S_0]$ and $(cc)_{\bf \bar{3}}[^3S_1]$ configurations together, we obtain the relative importance of different production mechanisms
\begin{eqnarray}
\left.\sigma_{g+g}^{p\rm Au}: \sigma_{g+c}^{p\rm Au}: \sigma_{c+c}^{p\rm Au}\right|_{\rm RHIC} \simeq 647 : 2720 : 1,  \\
\left.\sigma_{g+g}^{\rm AuAu}: \sigma_{g+c}^{\rm AuAu}: \sigma_{c+c}^{\rm AuAu}\right|_{\rm RHIC} \simeq 645 : 2735 : 1, \\
\left.\sigma_{g+g}^{p\rm Pb}: \sigma_{g+c}^{p\rm Pb}: \sigma_{c+c}^{p\rm Pb}\right|_{\rm LHC} \simeq 187 : 344 : 1,   \\
\left.\sigma_{g+g}^{\rm PbPb}: \sigma_{g+c}^{\rm PbPb}: \sigma_{c+c}^{\rm PbPb}\right|_{\rm LHC} \simeq 187 : 351 : 1.
\end{eqnarray}
Contributions from the $g+c$ mechanism are larger than the usually considered gluon-gluon fusion mechanism. As shall be shown below, this is caused by the fact that the cross-section for $g+c$ mechanism is larger than that of $g+g$ mechanism in small $p_t$ region.

\begin{table}[htb]
\begin{center}
\begin{tabular}{|c|c|c|c|c|}
\hline
- & \multicolumn{2}{|c|}{RHIC} & \multicolumn{2}{|c|} {LHC} \\
\hline
$\sqrt{S_{\rm NN}}$ (TeV) &$p$-Au(0.2) & Au-Au(0.2) & $p$-Pb(8.16) & Pb-Pb(5.02) \\
\hline
~$R$ ~& 6.13 & 6.19 & 3.39 & 3.44  \\
\hline
\end{tabular}
\caption{$R$ value which shows the importance of extrinsic charm mechanisms.}
\label{rrat}
\end{center}
\end{table}

To see how the extrinsic charm mechanism affects the $\Xi_{cc}$ production, we define a ratio
\begin{eqnarray}\label{Rate}
R = \frac{\sigma_{tot}}{\sigma_{gg \to \Xi_{cc}(cc)_{\bf \bar{3}}[^3S_1]}},
\end{eqnarray}
where $\sigma_{tot}$ stands for the summation of total cross sections for all the considered production mechanisms and diquark configurations, and $\sigma_{gg \to \Xi_{cc}(cc)_{\bf \bar{3}}[^3S_1]}$ is the cross section for usually considered gluon-gluon fusion via $gg \to \Xi_{cc}(cc)_{\bf \bar{3}}[^3S_1]$. The $R$ values are put in Table~\ref{rrat}. Table~\ref{rrat} shows that the extrinsic charm mechanism plays significant role via the $p$-N and the N-N collisions at the RHIC and the LHC, respectively.

\subsection{Differential distributions of the $\Xi_{cc}$ production via $p$-N and N-N collisions}

\begin{table*}[htb]
\begin{center}
\begin{tabular}{|c|c|c|c|c|c|c|}
\hline
- & \multicolumn{3}{|c|}{$p$-Au (0.2 TeV)} & \multicolumn{3}{|c|} {Au-Au (0.2 TeV)} \\
\hline
~-~ & $p_t>2\,\rm GeV$ & $p_t>4\,\rm GeV$ & $p_t>6\,\rm GeV$ & $p_t>2\,\rm GeV$ & $p_t>4\,\rm GeV$ & $p_t>6\,\rm GeV$ \\
\hline
~$\sigma(gg\to (cc)_{\bf 6}[^1S_0])$ ~& $4.79\times10^{-2}$ & $8.58\times10^{-3}$ & $1.36\times10^{-3}$ & $1.05\times10^{1}$ & 1.88 & $2.89\times10^{-1}$ \\
\hline
~$\sigma(gg\to (cc)_{\bar{\bf 3}}[^3S_1])$ ~& $2.21\times10^{-1}$ & $3.08\times10^{-2}$ & $4.12 \times10^{-3}$ & $4.91\times10^{1}$ & 6.74 & $8.68\times10^{-1}$ \\
\hline
~$\sigma(gc\to (cc)_{\bf 6}[^1S_0])$ ~& $7.17\times10^{-2}$ & $7.51\times10^{-3}$ & $8.93 \times10^{-4}$ & $1.70 \times10^{1}$ & 1.76 & $2.00\times10^{-1}$ \\
\hline
~$\sigma(gc\to (cc)_{\bar{\bf 3}}[^3S_1])$ ~& $6.12\times10^{-1}$ & $4.41\times10^{-2}$ & $4.33\times10^{-3}$ & $1.46\times10^{2}$ & $1.04\times10^{1}$ & $9.53\times10^{-1}$ \\
\hline
~$\sigma(cc\to (cc)_{\bf 6}[^1S_0])$ ~& $4.05\times10^{-5}$ & $4.05\times10^{-5}$ & $3.30\times10^{-5}$ & $8.96\times10^{-3}$ & $8.96\times10^{-3}$ & $7.24\times10^{-3}$ \\
\hline
~$\sigma(cc\to (cc)_{\bar{\bf 3}}[^3S_1])$ ~& $1.01\times10^{-3}$ & $1.01 \times10^{-3}$ & $8.64\times10^{-4}$ & $2.22\times10^{-1}$ & $2.22\times10^{-1}$ & $1.89\times10^{-1}$ \\
\hline
\end{tabular}
\caption{Total cross sections (in unit: $\mu$b) for the $\Xi_{cc}$ production via $g+g$, $g+c$, and $c+c$ mechanisms via $p$-Au and Au-Au collisions under various transverse momentum ($p_t$) cuts at the RHIC.}
\label{ptr}
\end{center}
\end{table*}

\begin{table*}
\begin{center}
\begin{tabular}{|c|c|c|c|c|c|c|}
\hline
- & \multicolumn{3}{|c|}{$p$-Pb (8.16 TeV)} & \multicolumn{3}{|c|} {Pb-Pb (5.02 TeV)} \\
\hline
~-~ & $p_t>2\,\rm GeV$ & $p_t>4\,\rm GeV$ & $p_t>6\,\rm GeV$ & $p_t>2\,\rm GeV$ & $p_t>4\,\rm GeV$ & $p_t>6\,\rm GeV$ \\
\hline
~$\sigma(gg\to (cc)_{\bf 6}[^1S_0])$ ~& 5.64 & 1.95 & $6.38\times10^{-1}$ & $6.39\times10^{2}$ & $2.20\times10^{2}$ & $7.08\times10^{1}$ \\
\hline
~$\sigma(gg\to (cc)_{\bar{\bf 3}}[^3S_1])$ ~& $2.64\times10^{1}$ & 7.81 & 2.27 & $2.99\times10^{3}$ & $8.86\times10^{2}$ & $2.53\times10^{2}$ \\
\hline
~$\sigma(gc\to (cc)_{\bf 6}[^1S_0])$ ~& 3.81 & $7.51\times10^{-1}$ & $1.72\times10^{-1}$ & $4.33\times10^{2}$ & $8.48\times10^{1}$ & $1.98\times10^{1}$ \\
\hline
~$\sigma(gc\to (cc)_{\bar{\bf 3}}[^3S_1])$ ~& $3.42\times10^{1}$ & 5.08 & 1.00 & $3.85\times10^{3}$ & $5.73\times10^{2}$ & $1.14\times10^{2}$ \\
\hline
~$\sigma(cc\to (cc)_{\bf 6}[^1S_0])$ ~& $1.05\times10^{-2}$ & $1.05\times10^{-2}$ & $9.46\times10^{-3}$ & 1.25 & 1.25 & 1.16\\
\hline
~$\sigma(cc\to (cc)_{\bar{\bf 3}}[^3S_1])$ ~& $2.94\times10^{-1}$ & $2.94\times10^{-1}$ & $2.67\times10^{-1}$ & $3.32\times10^{1}$ & $3.32\times10^{1}$ & $3.13\times10^{1}$ \\
\hline
\end{tabular}
\caption{Total cross sections (in unit: $\mu$b) for the $\Xi_{cc}$ production via $g+g$, $g+c$, and $c+c$ mechanisms via $p$-Pb and Pb-Pb collisions under various transverse momentum ($p_t$) cuts at the LHC.}
\label{ptl}
\end{center}
\end{table*}

\begin{table*}[htb]
\begin{center}
\begin{tabular}{|c|c|c|c|c|c|c|}
\hline
- & \multicolumn{3}{|c|}{$p$-Au (0.2 TeV)} & \multicolumn{3}{|c|} {Au-Au (0.2 TeV)} \\
\hline
$y_{\rm cut}$ & $|y|<1$ & $|y|<2$ & $|y|<3$ & $|y|<1$ & $|y|<2$ & $|y|<3$ \\
\hline
~$\sigma(gg\to (cc)_{\bf 6}[^1S_0])$ ~& $6.72\times10^{-2}$ & $9.94\times10^{-2}$ & $1.05\times10^{-1}$ & $1.54\times10^{1}$ & $2.22\times10^{1}$ & $2.30\times10^{1}$ \\
\hline
~$\sigma(gg\to (cc)_{\bar{\bf 3}}[^3S_1])$ ~& $3.72\times10^{-1}$ & $5.47\times10^{-1}$ & $5.75\times10^{-1}$ & $8.56\times10^{1}$ & $1.22\times10^{2}$ & $1.26\times10^{2}$ \\
\hline
~$\sigma(gc\to (cc)_{\bf 6}[^1S_0])$ ~& $1.42\times10^{-1}$ & $2.44\times10^{-1}$ & $2.79\times10^{-1}$ & $4.66\times10^{1}$ & $6.15\times10^{1}$ & $6.30\times10^{1}$ \\
\hline
~$\sigma(gc\to (cc)_{\bar{\bf 3}}[^3S_1])$ ~& 1.23 & 2.21 & 2.54 & $4.28\times10^{2}$ & $5.61\times10^{2}$ & $5.75\times10^{2}$ \\
\hline
~$\sigma(cc\to (cc)_{\bf 6}[^1S_0])$ ~& $3.16\times10^{-5}$ & $4.02\times10^{-5}$ & $4.05\times10^{-5}$ & $7.45\times10^{-3}$ & $8.95\times10^{-3}$ & $8.96\times10^{-3}$ \\
\hline
~$\sigma(cc\to (cc)_{\bar{\bf 3}}[^3S_1])$ ~& $7.88\times10^{-4}$ & $1.00\times10^{-3}$ & $1.01\times10^{-3}$ & $1.84\times10^{-1}$ & $2.22\times10^{-1}$ & $2.22\times10^{-1}$ \\
\hline
\end{tabular}
\caption{Total cross sections (in unit: $\mu$b) for the $\Xi_{cc}$ production via $g+g$, $g+c$, and $c+c$ mechanisms via $p$-Au and Au-Au collisions under various rapidity ($y$) cuts at the RHIC.}
\label{yr}
\end{center}
\end{table*}

\begin{table*}[htb]
\begin{center}
\begin{tabular}{|c|c|c|c|c|c|c|}
\hline
- & \multicolumn{3}{|c|}{$p$-Pb (8.16 TeV)} & \multicolumn{3}{|c|} {Pb-Pb (5.02 TeV)} \\
\hline
$y_{\rm cut}$ & $|y|<1$ & $|y|<2$ & $|y|<3$ & $|y|<1$ & $|y|<2$ & $|y|<3$ \\
\hline
~$\sigma(gg\to (cc)_{\bf 6}[^1S_0])$ ~& 2.53 & 4.96 & 6.96 & $2.99\times10^{2}$ & $5.83\times10^{2}$ & $8.20\times10^{2}$ \\
\hline
~$\sigma(gg\to (cc)_{\bar{\bf 3}}[^3S_1])$ ~& $1.33\times10^{1}$ & $2.59\times10^{1}$ & $3.62\times10^{1}$ & $1.54\times10^{3}$ & $3.02\times10^{3}$ & $4.28\times10^{3}$ \\
\hline
~$\sigma(gc\to (cc)_{\bf 6}[^1S_0])$ ~& 5.17 & 7.51 & 9.36 & $4.49\times10^{2}$ & $7.25\times10^{2}$ & $9.42\times10^{2}$ \\
\hline
~$\sigma(gc\to (cc)_{\bar{\bf 3}}[^3S_1])$ ~& $3.30\times10^{1}$ & $5.36\times10^{1}$ & $7.06\times10^{1}$ & $4.10\times10^{3}$ & $6.53\times10^{3}$ & $8.52\times10^{3}$ \\
\hline
~$\sigma(cc\to (cc)_{\bf 6}[^1S_0])$ ~& $4.54\times10^{-3}$ & $8.39\times10^{-3}$ & $9.95\times10^{-3}$ & $5.86\times10^{-1}$ & 1.07 & 1.22 \\
\hline
~$\sigma(cc\to (cc)_{\bar{\bf 3}}[^3S_1])$ ~& $1.21\times10^{-1}$ & $2.35\times10^{-1}$ & $2.83\times10^{-1}$ & $1.56\times10^{1}$ & $2.86\times10^{1}$ & $3.25\times10^{1}$ \\
\hline
\end{tabular}
\caption{Total cross sections (in unit: $\mu$b) for the $\Xi_{cc}$ production via $g+g$, $g+c$, and $c+c$ mechanisms via $p$-Pb and Pb-Pb collisions under various rapidity ($y$) cuts at the LHC.}
\label{yl}
\end{center}
\end{table*}

At the hadronic colliders, the small $p_t$ and/or large rapidity $y$ (the produced baryons move very close to the beam direction) cannot be detected by the detectors directly, and such kind of events cannot be utilized for experimental studies in common cases. In this subsection, we perform detailed calculations and discussions under different $p_t$ cuts and $y$ cuts.

We adopt three typical $p_t$ cuts, $p_t>2$ GeV, $p_t>4$ GeV and $p_t>6$ GeV, to show how the production cross sections change with the $\Xi_{cc}$ transverse momentum. The results are presented in Tables~\ref{ptr} and \ref{ptl}. Table~\ref{ptr} shows that the total cross sections for the $\Xi_{cc}$ via $p$-Au and Au-Au collisions at the RHIC shall be reduced by about $54\% \sim 99\%$ when the $p_t$-cut varying from $2$ GeV to $6$ GeV. Similarly, the production rates via $p$-Pb and Pb-Pb collisions shall be reduced by about $38\% \sim 98\%$. This shows that small $p_t$ region provides significant contributions to the $\Xi_{cc}$ production via $p$-N and N-N collisions at the RHIC and the LHC.

We adopt three typical rapidity cuts, $|y|<1$, $|y|<2$, and $|y|<3$,  to show how the production cross sections change with the $\Xi_{cc}$ rapidity. The results are presented in Tables~\ref{yr} and \ref{yl}. Table~\ref{yr} shows that at the RHIC, the $\Xi_{cc}$ events mainly distribute in the rapidity region of $y \in [-3,3]$; At the LHC, the $\Xi_{cc}$ events distribute within a broader rapidity range.

\begin{figure}[htb]
\includegraphics[width=0.5\textwidth]{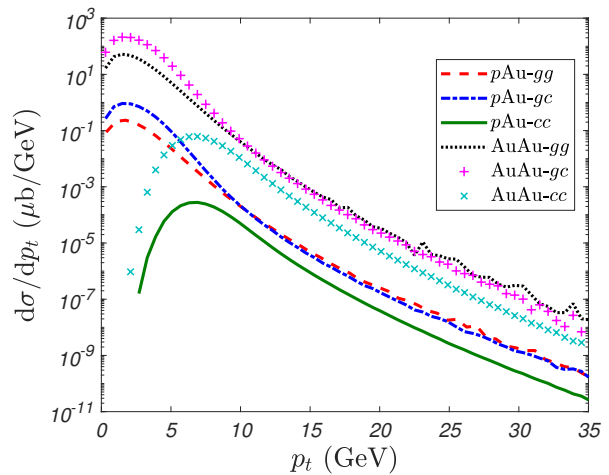}
\caption{The $p_t$-distributions of the $\Xi_{cc}$ production via $g+g$, $g+c$, and $c+c$ mechanisms via $p$-Au and Au-Au collisions  with $\sqrt{S_{p\rm Au}}=0.2$ TeV and $\sqrt{S_{\rm AuAu}}=0.2$ TeV at the RHIC. }  \label{ptdr}
\end{figure}

\begin{figure}[htb]
\includegraphics[width=0.5\textwidth]{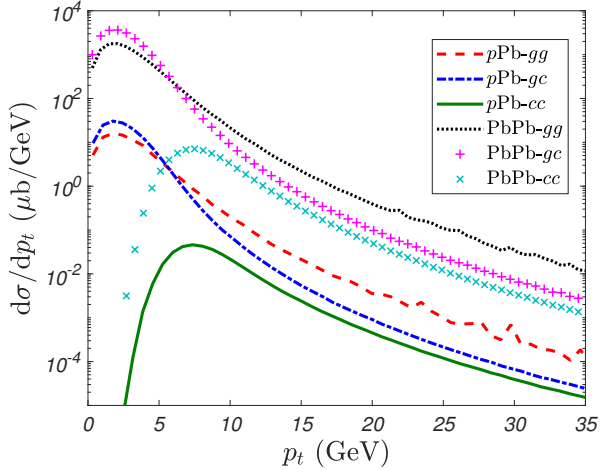}
\caption{The $p_t$-distribution of the $\Xi_{cc}$ production via $g+g$, $g+c$, and $c+c$ mechanisms via $p$-Pb and Pb-Pb collisions with $\sqrt{S_{p\rm Pb}}=8.16$ TeV and $\sqrt{S_{\rm PbPb}}=5.02$ TeV at the LHC.} \label{ptdl}
\end{figure}

We present the $\Xi_{cc}$ $p_t$ distributions via $p$-N and N-N collisions at the RHIC and the LHC in Figs.~\ref{ptdr},  \ref{ptdl}. For each production mechanism, contributions from $(cc)_{\bf 6}[^1S_0]$ and $(cc)_{\bf \bar{3}}[^3S_1]$ configurations have been summed for convenience. The $p_t$ distributions at the RHIC and LHC have similar shapes, which shall first increase with the increment of $p_t$ in small $p_t$ region and then decrease quickly in large $p_t$ region. In small $p_t$ region, the $g+c$ mechanism is larger than $g+g$ mechanism via $p$-N and N-N collisions. If the experimental measurements can be extended to small $p_t$ region, then one may study extrinsic charm mechanism by measuring the $\Xi_{cc}$ events.

\begin{figure}[htb]
\includegraphics[width=0.5\textwidth]{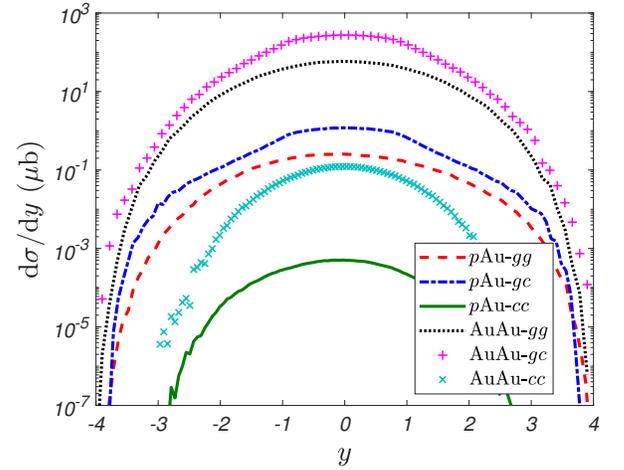}
\caption{The $y$-distributions of the $\Xi_{cc}$ production via $g+g$, $g+c$, and $c+c$ mechanisms via $p$-Au and Au-Au collisions  with $\sqrt{S_{p\rm Au}}=0.2$ TeV and $\sqrt{S_{\rm AuAu}}=0.2$ TeV at the RHIC.}  \label{rapR}
\end{figure}

\begin{figure}[htb]
\includegraphics[width=0.5\textwidth]{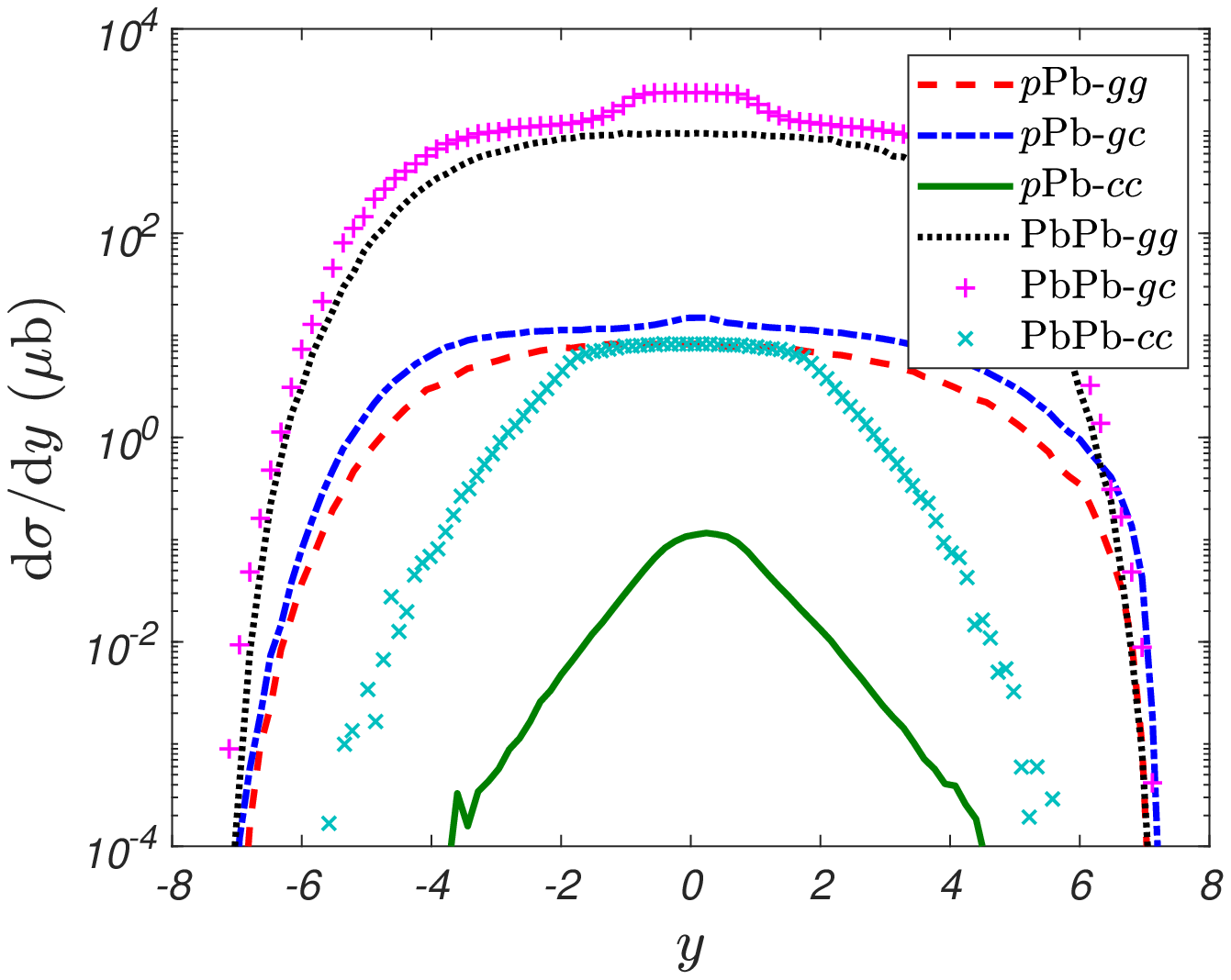}
\caption{The $y$-distributions of the $\Xi_{cc}$ production via $g+g$, $g+c$, and $c+c$ mechanisms via $p$-Pb and Pb-Pb collisions with $\sqrt{S_{p\rm Pb}}=8.16$ TeV and $\sqrt{S_{\rm PbPb}}=5.02$ TeV at the LHC. }  \label{rapL}
\end{figure}

\begin{figure}[htb]
\includegraphics[width=0.5\textwidth]{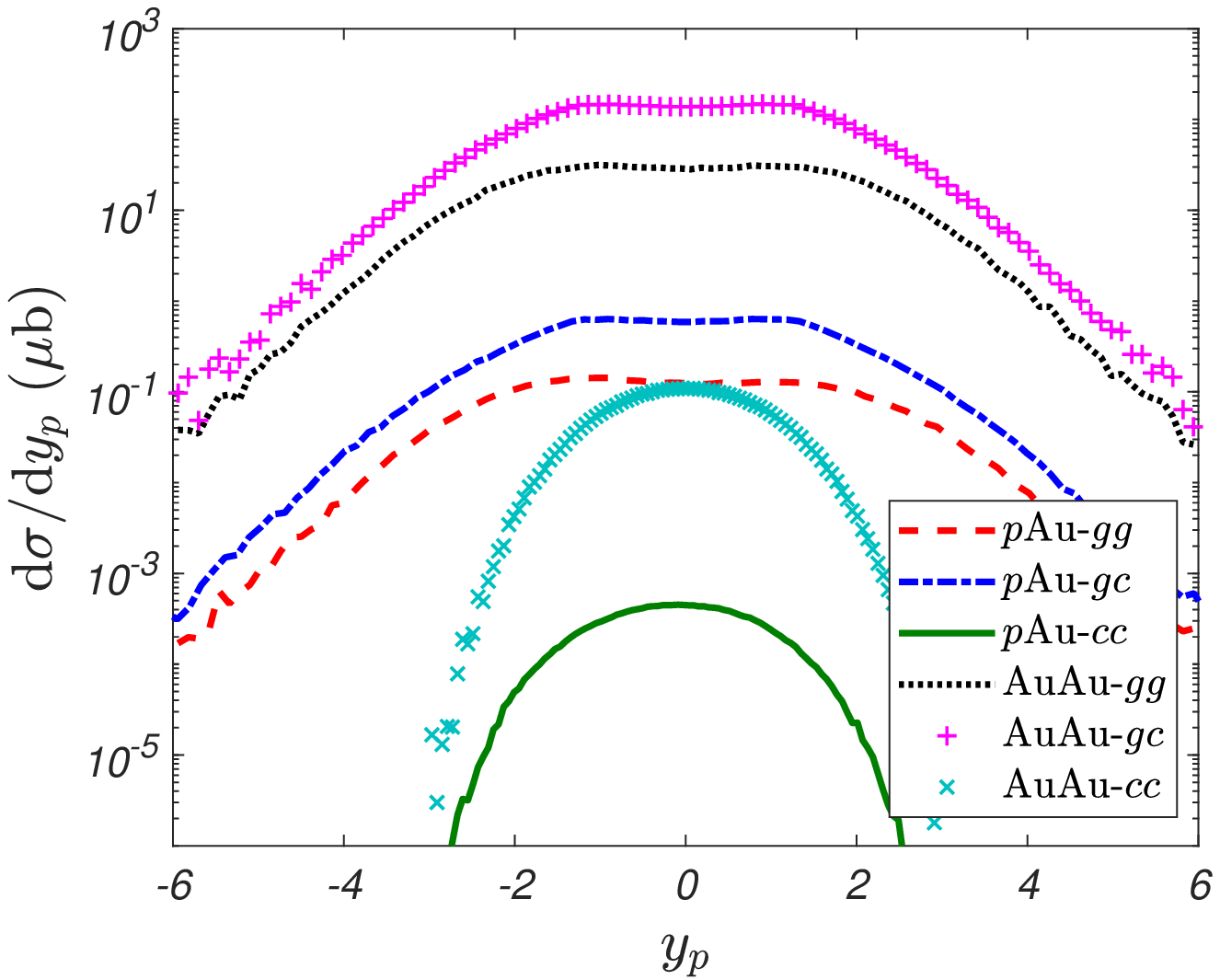}
\caption{The pseudo-rapidity ($y_p$)-distributions of the $\Xi_{cc}$ production via $g+g$, $g+c$, and $c+c$ mechanisms via $p$-Au and Au-Au collisions  with $\sqrt{S_{p\rm Au}}=0.2$ TeV and $\sqrt{S_{\rm AuAu}}=0.2$ TeV at the RHIC.}  \label{prapR}
\end{figure}

\begin{figure}[htb]
\includegraphics[width=0.5\textwidth]{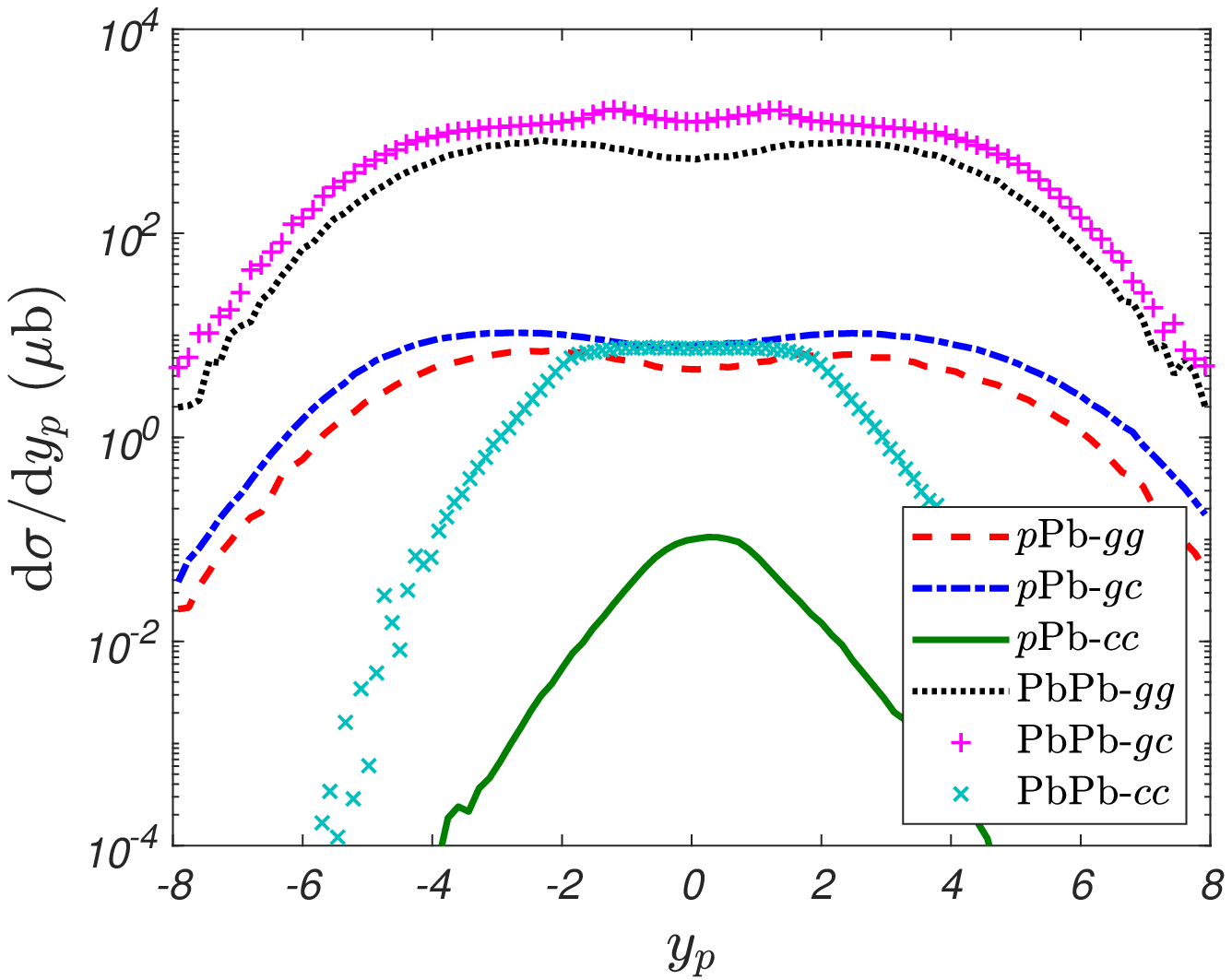}
\caption{The pseudo-rapidity ($y_p$)-distributions of the $\Xi_{cc}$ production via $g+g$, $g+c$, and $c+c$ mechanisms via $p$-Pb and Pb-Pb collisions with $\sqrt{S_{p\rm Pb}}=8.16$ TeV and $\sqrt{S_{\rm PbPb}}=5.02$ TeV at the LHC. }  \label{prapL}
\end{figure}

We present the rapidity ($y$) and pseudo-rapidity ($y_p$) distributions of the $\Xi_{cc}$ production at the RHIC and LHC in Figs. \ref{rapR}, \ref{rapL}, \ref{prapR}, and \ref{prapL}, respectively. For the dominant $g+g$ and $g+c$ mechanisms, there are plateaus for $|y| \le 2$ or $|y_p| \le 2$ at the RHIC, which become broader at the LHC, i.e. $|y| \le 5$ and $|y_p| \le 5$.

\subsection{Theoretical uncertainties for $\Xi_{cc}$ production}

\begin{table*}[htb]
\begin{center}
\begin{tabular}{|c|c|c|c|c|c|c|}
\hline
- & \multicolumn{3}{|c|}{$p$-Au (0.2 TeV)} & \multicolumn{3}{|c|} {Au-Au (0.2 TeV)} \\
\hline
$m_c$ & 1.65 GeV & 1.75 GeV & 1.85 GeV & 1.65 GeV & 1.75 GeV & 1.85 GeV  \\
\hline
~$\sigma(gg\to (cc)_{\bf 6}[^1S_0])$ ~& $1.72\times10^{-1}$ & $1.05\times10^{-1}$ & $6.53\times10^{-2}$ & $3.76\times10^{1}$ & $2.30\times10^{1}$ & $1.44\times10^{1}$ \\
\hline
~$\sigma(gg\to (cc)_{\bar{\bf 3}}[^3S_1])$ ~& $9.52\times10^{-1}$ & $5.79\times10^{-1}$ & $3.60\times10^{-1}$ & $2.07\times10^{2}$ & $1.27\times10^{2}$ & $7.91\times10^{1}$ \\
\hline
~$\sigma(gc\to (cc)_{\bf 6}[^1S_0])$ ~& $4.07\times10^{-1}$ & $2.83\times10^{-1}$ & $1.98\times10^{-1}$ & $8.70\times10^{1}$ & $6.26\times10^{1}$ & $4.58\times10^{1}$ \\
\hline
~$\sigma(gc\to (cc)_{\bar{\bf 3}}[^3S_1])$ ~& 3.70 & 2.58 & 1.80 & $7.89\times10^{2}$ & $5.73\times10^{2}$ & $4.15\times10^{2}$ \\
\hline
~$\sigma(cc\to (cc)_{\bf 6}[^1S_0])$ ~& $7.63\times10^{-5}$ & $4.06\times10^{-5}$ & $2.88\times10^{-5}$ & $1.72\times10^{-2}$ & $9.00\times10^{-3}$ & $6.35\times10^{-3}$ \\
\hline
~$\sigma(cc\to (cc)_{\bar{\bf 3}}[^3S_1])$ ~& $1.86\times10^{-3}$ & $1.01\times10^{-3}$ & $7.01\times10^{-4}$ & $4.15\times10^{-1}$ & $2.23\times10^{-1}$ & $1.54\times10^{-1}$ \\
\hline
\end{tabular}
\caption{Total cross sections (in unit: $\mu$b) of $\Xi_{cc}$ production via $g+g$, $g+c$, and $c+c$ mechanisms for three typical $c$-quark masses via $p$-Au and Au-Au collisions at the RHIC.}
\label{mcr}
\end{center}
\end{table*}

\begin{table*}[htb]
\begin{center}
\begin{tabular}{|c|c|c|c|c|c|c|}
\hline
- & \multicolumn{3}{|c|}{$p$-Pb (8.16 TeV)} & \multicolumn{3}{|c|} {Pb-Pb (5.02 TeV)} \\
\hline
$m_c$ & 1.65 GeV & 1.75 GeV & 1.85 GeV & 1.65 GeV & 1.75 GeV & 1.85 GeV  \\
\hline
~$\sigma(gg\to (cc)_{\bf 6}[^1S_0])$ ~& $1.29\times10^{1}$ & 9.20 & 6.58 & $1.46\times10^{3}$ & $1.04\times10^{3}$ & $7.52\times10^{2}$ \\
\hline
~$\sigma(gg\to (cc)_{\bar{\bf 3}}[^3S_1])$ ~& $6.71\times10^{1}$ & $4.79\times10^{1}$ & $3.46\times10^{1}$ & $7.61\times10^{3}$ & $5.38\times10^{3}$ & $3.92\times10^{3}$ \\
\hline
~$\sigma(gc\to (cc)_{\bf 6}[^1S_0])$ ~& $1.48\times10^{1}$ & $1.17\times10^{1}$ & $9.40$ & $1.50\times10^{3}$ & $1.19\times10^{3}$ & $9.52\times10^{2}$ \\
\hline
~$\sigma(gc\to (cc)_{\bar{\bf 3}}[^3S_1])$ ~& $1.18\times10^{2}$ & $9.31\times10^{1}$ & $7.44\times10^{1}$ & $1.38\times10^{4}$ & $1.09\times10^{4}$ & $8.67\times10^{3}$\\
\hline
~$\sigma(cc\to (cc)_{\bf 6}[^1S_0])$ ~& $1.60\times10^{-2}$ & $1.05\times10^{-2}$ & $8.24\times10^{-3}$ & 1.90 & 1.25 & $9.81\times10^{-1}$ \\
\hline
~$\sigma(cc\to (cc)_{\bar{\bf 3}}[^3S_1])$ ~& $4.39\times10^{-1}$ & $2.94\times10^{-1}$ & $2.28\times10^{-1}$ & $4.96\times10^{1}$ & $3.32\times10^{1}$ & $2.56\times10^{1}$\\
\hline
\end{tabular}
\caption{Total cross sections (in unit: $\mu$b) of $\Xi_{cc}$ production via $g+g$, $g+c$, and $c+c$ mechanisms for three typical $c$-quark masses via $p$-Pb and Pb-Pb collisions at the LHC. }
\label{mcl}
\end{center}
\end{table*}

The non-perturbative matrix elements are overall parameters, thus their uncertainties can be conveniently suppressed when we know their exact values. In this subsection, we discuss the other two important uncertainties, which are from the choices of charm quark mass and renormalization scale. For the purpose, we take  $m_c=1.75 \pm 0.1\,\rm GeV$ and three frequently used scales, e.g. the transverse mass $M_t$ of $\Xi_{cc}$, $\sqrt{\hat{s}}$ and $\sqrt{\hat{s}}/2$, where $\sqrt{\hat{s}}$ is the center-of-mass energy of the corresponding subprocess, to do the calculation. For convenience, when considering one parameter's uncertainty, the other parameters shall be kept to their central values.

Total cross sections of $\Xi_{cc}$ production for $m_c=1.75 \pm 0.1\,\rm GeV$ are presented in Tables~\ref{mcr} and \ref{mcl}, which are for RHIC and LHC, respectively. Tables~\ref{mcr} and \ref{mcl} indicate the total cross sections decrease with the increment of $m_c$. The uncertainties at the RHIC via $p$-Au and Au-Au collisions with $\sqrt{S_{p{\rm Au}}}=\sqrt{S_{{\rm Au Au}}}=0.2{\rm TeV}$ are
\begin{eqnarray}
\sigma_{gg\to (cc)_{\bf 6}[^1S_0]}^{p\rm Au} &=& \left(1.05^{+0.67}_{-0.397}\right)\times10^{-1}\;{\rm \mu b}, \\
\sigma_{gg\to (cc)_{\bar{\bf 3}}[^3S_1]}^{p\rm Au} &=& \left(5.79^{+3.73}_{-2.19}\right)\times10^{-1}\;{\rm \mu b}, \\
\sigma_{gc\to (cc)_{\bf 6}[^1S_0]}^{p\rm Au} &=& \left(2.83^{+1.24}_{-0.85}\right)\times10^{-1}\;{\rm \mu b}, \\
\sigma_{gc\to (cc)_{\bar{\bf 3}}[^3S_1]}^{p\rm Au} &=& 2.58^{+1.12}_{-0.78}\;{\rm \mu b}, \\
\sigma_{cc\to (cc)_{\bf 6}[^1S_0]}^{p\rm Au} &=& \left(4.06^{+3.57}_{-1.18}\right)\times10^{-5}\;{\rm \mu b}, \\
\sigma_{cc\to (cc)_{\bar{\bf 3}}[^3S_1]}^{p\rm Au} &=& \left(1.01^{+0.85}_{-0.309}\right)\times10^{-3}\;{\rm \mu b}, \\
\sigma_{gg\to (cc)_{\bf 6}[^1S_0]}^{\rm AuAu} &=& \left(2.30^{+1.46}_{-0.86}\right)\times10^{1}\;{\rm \mu b}, \\
\sigma_{gg\to (cc)_{\bar{\bf 3}}[^3S_1]}^{\rm AuAu} &=& \left(1.27^{+0.800}_{-0.479}\right)\times10^{2}\;{\rm \mu b}, \\
\sigma_{gc\to (cc)_{\bf 6}[^1S_0]}^{\rm AuAu} &=& \left(6.26^{+2.44}_{-1.68}\right)\times10^1\;{\rm \mu b}, \\
\sigma_{gc\to (cc)_{\bar{\bf 3}}[^3S_1]}^{\rm AuAu} &=& \left(5.73^{+2.16}_{-1.58}\right)\times10^{2}\;{\rm \mu b}, \\
\sigma_{cc\to (cc)_{\bf 6}[^1S_0]}^{\rm AuAu} &=& \left(9.00^{+8.20}_{-2.65}\right)\times10^{-3}\;{\rm \mu b}, \\
\sigma_{cc\to (cc)_{\bar{\bf 3}}[^3S_1]}^{\rm AuAu} &=& \left(2.23^{+1.92}_{-0.69}\right)\times10^{-1}\;{\rm \mu b}.
\end{eqnarray}
The uncertainties at the LHC via $p$-Pb and Pb-Pb collisions with $\sqrt{S_{p{\rm Pb}}}=8.16$ TeV and $\sqrt{S_{{\rm Pb Pb}}}=5.02$ TeV are
\begin{eqnarray}
\sigma_{gg\to (cc)_{\bf 6}[^1S_0]}^{p\rm Pb} &=& 9.20^{+3.70}_{-2.62}\;{\rm \mu b}, \\
\sigma_{gg\to (cc)_{\bar{\bf 3}}[^3S_1]}^{p\rm Pb} &=& \left(4.79^{+1.92}_{-1.33}\right)\times10^1\;{\rm \mu b}, \\
\sigma_{gc\to (cc)_{\bf 6}[^1S_0]}^{p\rm Pb} &=& \left(1.17^{+0.310}_{-0.230}\right)\times10^1\;{\rm \mu b}, \\
\sigma_{gc\to (cc)_{\bar{\bf 3}}[^3S_1]}^{p\rm Pb} &=& \left(9.31^{+2.49}_{-1.87}\right)\times10^1\;{\rm \mu b}, \\
\sigma_{cc\to (cc)_{\bf 6}[^1S_0]}^{p\rm Pb} &=& \left(1.05^{+0.550}_{-0.226}\right)\times10^{-2}\;{\rm \mu b}, \\
\sigma_{cc\to (cc)_{\bar{\bf 3}}[^3S_1]}^{p\rm Pb} &=& \left(2.94^{+1.45}_{-0.66}\right) \times10^{-1}\;{\rm \mu b}, \\
\sigma_{gg\to (cc)_{\bf 6}[^1S_0]}^{\rm PbPb} &=& \left(1.04^{+0.420}_{-0.288}\right)\times10^{3}\;{\rm \mu b}, \\
\sigma_{gg\to (cc)_{\bar{\bf 3}}[^3S_1]}^{\rm PbPb} &=& \left(5.38^{+2.23}_{-1.46}\right)\times10^{3}\;{\rm \mu b}, \\
\sigma_{gc\to (cc)_{\bf 6}[^1S_0]}^{\rm PbPb} &=& \left(1.19^{+0.310}_{-0.238}\right)\times10^{3}\;{\rm \mu b}, \\
\sigma_{gc\to (cc)_{\bar{\bf 3}}[^3S_1]}^{\rm PbPb} &=& \left(1.09^{+0.290}_{-0.223}\right)\times10^{4}\;{\rm \mu b}, \\
\sigma_{cc\to (cc)_{\bf 6}[^1S_0]}^{\rm PbPb} &=& 1.25^{+0.650}_{-0.269}\;{\rm \mu b}, \\
\sigma_{cc\to (cc)_{\bar{\bf 3}}[^3S_1]}^{\rm PbPb} &=& \left(3.33^{+1.64}_{-0.76}\right)\times10^1\;{\rm \mu b}.
\end{eqnarray}

\begin{table*}[htb]
\begin{center}
\begin{tabular}{|c|c|c|c|c|c|c|}
\hline
- & \multicolumn{3}{|c|}{$p$-Au (0.2 TeV)} & \multicolumn{3}{|c|} {Au-Au (0.2 TeV)} \\
\hline
~$\mu_R$~ & $\sqrt{\hat{s}}$ & $\sqrt{\hat{s}}/2$ & $M_t$ & $\sqrt{\hat{s}}$ & $\sqrt{\hat{s}}/2$ & $M_t$ \\
\hline
~$\sigma(gg\to (cc)_{\bf 6}[^1S_0])$ ~& $2.71\times10^{-2}$ & $5.98\times10^{-2}$ & $1.05\times10^{-1}$ & 5.85 & $1.30\times10^{1}$ & $2.30\times10^{1}$ \\
\hline
~$\sigma(gg\to (cc)_{\bar{\bf 3}}[^3S_1])$ ~& $1.42\times10^{-1}$ & $3.15\times10^{-1}$ & $5.79\times10^{-1}$ & $3.09\times10^{1}$ & $6.85\times10^{1}$ & $1.27\times10^{2}$ \\
\hline
~$\sigma(gc\to (cc)_{\bf 6}[^1S_0])$ ~& $2.09\times10^{-1}$ & $2.86\times10^{-1}$ & $2.83\times10^{-1}$ & $4.21\times10^{1}$ & $6.11\times10^{1}$ & $6.26\times10^{1}$ \\
\hline
~$\sigma(gc\to (cc)_{\bar{\bf 3}}[^3S_1])$ ~& 1.89 & 2.58 & 2.58 & $3.83\times10^{2}$ & $5.56\times10^{2}$ & $5.73\times10^{2}$ \\
\hline
~$\sigma(cc\to (cc)_{\bf 6}[^1S_0])$ ~& $3.11\times10^{-5}$ & $4.00\times10^{-5}$ & $4.06\times10^{-5}$ & $6.81\times10^{-3}$ & $8.91\times10^{-3}$ & $9.00\times10^{-3}$ \\
\hline
~$\sigma(cc\to (cc)_{\bar{\bf 3}}[^3S_1])$ ~& $7.70\times10^{-4}$ & $9.98\times10^{-4}$ & $1.01\times10^{-3}$ & $1.68\times10^{-1}$ & $2.21\times10^{-1}$ & $2.23\times10^{-1}$ \\
\hline
\end{tabular}
\caption{Total cross sections (in unit: $\mu$b) of $\Xi_{cc}$ production via $g+g$, $g+c$, and $c+c$ mechanisms for three typical renormalization scale $\mu_R$ via $p$-Au and Au-Au collisions at the RHIC. }   \label{sr}
\end{center}
\end{table*}

\begin{table*}
\begin{center}
\begin{tabular}{|c|c|c|c|c|c|c|}
\hline
- & \multicolumn{3}{|c|}{$p$-Pb (8.16 TeV)} & \multicolumn{3}{|c|} {Pb-Pb (5.02 TeV)} \\
\hline
~$\mu_R$~ & $\sqrt{\hat{s}}$ & $\sqrt{\hat{s}}/2$ & $M_t$ & $\sqrt{\hat{s}}$ & $\sqrt{\hat{s}}/2$ & $M_t$ \\
\hline
~$\sigma(gg\to (cc)_{\bf 6}[^1S_0])$ ~& 6.07 & 7.50 & 9.20 & $6.70\times10^{2}$ & $8.37\times10^{2}$ & $1.04\times10^{3}$ \\
\hline
~$\sigma(gg\to (cc)_{\bar{\bf 3}}[^3S_1])$ ~& $3.17\times10^{1}$ & $3.91\times10^{1}$ & $4.79\times10^{1}$ & $3.48\times10^{3}$ & $4.38\times10^{3}$ & $5.38\times10^{3}$ \\
\hline
~$\sigma(gc\to (cc)_{\bf 6}[^1S_0])$ ~& $1.12\times10^{1}$ & $1.10\times10^{1}$ & $1.17\times10^{1}$ & $1.16\times10^{3}$ & $1.11\times10^{3}$ & $1.19\times10^{3}$ \\
\hline
~$\sigma(gc\to (cc)_{\bar{\bf 3}}[^3S_1])$ ~& $9.18\times10^{1}$ & $8.76\times10^{1}$ & $9.31\times10^{1}$ & $1.06\times10^{4}$ & $1.02\times10^{4}$ & $1.09\times10^{4}$ \\
\hline
~$\sigma(cc\to (cc)_{\bf 6}[^1S_0])$ ~& $1.11\times10^{-2}$ & $1.05\times10^{-2}$ & $1.05\times10^{-2}$ & 1.31 & 1.25 & 1.25 \\
\hline
~$\sigma(cc\to (cc)_{\bar{\bf 3}}[^3S_1])$ ~& $3.13\times10^{-1}$ & $2.96\times10^{-1}$ & $2.94\times10^{-1}$ & $3.48\times10^{1}$ & $3.33\times10^{1}$ & $3.33\times10^{1}$ \\
\hline
\end{tabular}
\caption{Total cross sections (in unit: $\mu$b) of $\Xi_{cc}$ production via $g+g$, $g+c$, and $c+c$ mechanisms for three typical renormalization scale $\mu_R$ via $p$-Pb and Pb-Pb collisions at the LHC. }
\label{sl}
\end{center}
\end{table*}

The renormalization scale-setting problem is an important problem of fixed-order pQCD predictions~\cite{Wu:2013ei}. As an quantitative estimation of renormalization scale dependence, we choose three usually adopted values as the renormalization scale, i.e. the transverse mass $M_t$ of $\Xi_{cc}$, $\sqrt{\hat{s}}$ and $\sqrt{\hat{s}}/2$, where $\sqrt{\hat{s}}$ is the center-of-mass energy of the subprocess. Numerical results are presented in Tables~\ref{sr} and \ref{sl}. The scale uncertainties at the RHIC are large, which varies from $23\%$ to $76\%$ for various mechanisms via $p$-Au and Au-Au collisions, accordingly; while the scale uncertainty at the LHC is smaller, which varies from $4\%$ to $36\%$ for various mechanisms via $p$-Pb and Pb-Pb collisions. Thus we need a next-to-leading order calculation to achieve more accurate predictions, especially, by applying the principle of maximum conformality scale-setting approach~\cite{Brodsky:2012rj, Mojaza:2012mf}, the renormalization scale uncertainties can be eliminated.

\subsection{A simple discussion of $\Xi_{bc}$ and $\Xi_{bb}$ production at the RHIC and LHC}

In this subsection, we present a simple discussion of $\Xi_{bc}$ and $\Xi_{bb}$ production properties via $p$-N and N-N collisions at the RHIC and LHC. Their production mechanisms can be treated via the same way as those of $\Xi_{cc}$ production, and we adopt the generator GENXICC to do the calculation.

As for the input parameters, we take: $|\Psi_{bc}(0)|^2=0.065$ GeV$^3$ and $|\Psi_{bb}(0)|^2 = 0.152$ GeV$^3$~\cite{Baranov:1995rc}, and $M_{\Xi_{bc}}=6.9$ with $m_c=1.8$ GeV and $m_b=5.1$ GeV, $M_{\Xi_{bb}}=10.2$ GeV with $m_b=M_{\Xi_{bb}}/2$. And we set the renormalization scale as $M_t$. In different to the $\Xi_{cc}$ production, for the present case of $\Xi_{bc}$ and $\Xi_{bb}$, the extrinsic mechanisms shall be highly suppressed by the much smaller bottom-quark PDF, thus we shall only consider the dominant gluon-gluon fusion mechanism.

\begin{table*}[htb]
\begin{center}
\begin{tabular}{|c|c|c|c|c|}
\hline
- & \multicolumn{2}{|c|}{RHIC} & \multicolumn{2}{|c|}{LHC} \\
\hline
$\sqrt{S_{\rm NN}}$ (TeV)  & $p$-Au (0.2) & Au-Au (0.2) &  $p$-Pb (8.16)& Pb-Pb (5.02)  \\
\hline
$\sigma(gg\to (bc)_{\bf 6}[^1S_0])$  & $1.53\times10^{-3}$ & $3.27\times10^{-1}$  & $7.16\times10^{-1}$ & $7.97\times10^{1}$ \\
\hline
$\sigma(gg\to (bc)_{\bar{\bf 3}}[^3S_1])$  & $5.98\times10^{-3}$ & 1.28  & 2.92 & $3.28\times10^{2}$  \\
\hline
$\sigma(gg\to (bc)_{\bf 6}[^3S_1])$  & $1.06\times10^{-2}$ & 2.28  & 4.32  & $4.83\times10^{2}$  \\
\hline
$\sigma(gg\to (bc)_{\bar{\bf 3}}[^1S_0])$  & $1.65\times10^{-3}$ & $3.52\times10^{-1}$  & $7.70\times10^{-1}$ & $8.51\times10^{1}$  \\
\hline
$\sigma^{tot}(\Xi_{bc})$ & $1.98\times10^{-2}$ & 4.24  & 8.73 & $9.75\times10^{2}$  \\
\hline
$\sigma(gg\to (bb)_{\bf 6}[^1S_0])$  & $1.94\times10^{-5}$ & $3.71\times10^{-3}$ & $4.07\times10^{-2}$ & 4.40  \\
\hline
$\sigma(gg\to (bb)_{\bar{\bf 3}}[^3S_1])$  & $1.05\times10^{-4}$ & $2.01\times10^{-2}$  & $2.15\times10^{-1}$ & $2.35\times10^{1}$   \\
\hline
$\sigma^{tot}(\Xi_{bb})$  & $1.25\times10^{-4}$ & $2.38\times10^{-2}$  & $2.55\times10^{-1}$ & $2.79\times10^{1}$  \\
\hline
\end{tabular}
\caption{Total cross sections (in unit $\mu$b) for doubly heavy baryons $\Xi_{bc}$ and $\Xi_{bb}$ via the dominant gluon-gluon fusion mechanism via $p$-N, and N-N collisions at the RHIC and the LHC.}
\label{totcrobb}
\end{center}
\end{table*}

We present the total cross sections for $\Xi_{bc}$ and $\Xi_{bb}$ produced via $p$-N and N-N collisions at the RHIC and the LHC in Table~\ref{totcrobb}. By summing up different spin-color diquark configurations, we obtain
\begin{eqnarray}
\left.\sigma^{\rm tot}_{p\rm Au}(\Xi_{bc})\right|_{\rm RHIC} &=& 1.98\times10^{-2}\;{\rm \mu b}, \\
\left.\sigma^{\rm tot}_{p\rm Au}(\Xi_{bb})\right|_{\rm RHIC} &=& 1.25\times10^{-4}\;{\rm \mu b},  \\
\left.\sigma^{\rm tot}_{\rm AuAu}(\Xi_{bc})\right|_{\rm RHIC} &=& 4.24\;{\rm \mu b},  \\
\left.\sigma^{\rm tot}_{\rm AuAu}(\Xi_{bb})\right|_{\rm RHIC} &=& 2.38\times10^{-2}\;{\rm \mu b},  \\
\left.\sigma^{\rm tot}_{p\rm Pb}(\Xi_{bc})\right|_{\rm LHC} &=& 8.73\;{\rm \mu b}, \\
\left.\sigma^{\rm tot}_{p\rm Pb}(\Xi_{bb})\right|_{\rm LHC} &=& 2.55\times10^{-1}\;{\rm \mu b}, \\
\left.\sigma^{\rm tot}_{\rm PbPb}(\Xi_{bc})\right|_{\rm LHC} &=& 9.75\times10^{2}\;{\rm \mu b}, \\
\left.\sigma^{\rm tot}_{\rm PbPb}(\Xi_{bb})\right|_{\rm LHC} &=& 2.79\times10^{1}\;{\rm \mu b}.
\end{eqnarray}

To estimate the event numbers for $\Xi_{bc}$ and $\Xi_{bb}$ porduction, we adopt the same luminosities, as mentioned above, for the $p$-N and N-N collisions at the RHIC and LHC. Our results show that at the RHIC, $8.9 \times 10^4$ $\Xi_{bc}$ and $5.6 \times 10^2$ $\Xi_{bb}$ events can be generated via $p$-Au collision, $3.4 \times 10^5$ $\Xi_{bc}$ and $1.9 \times 10^3$ $\Xi_{bb}$ events via Au-Au collision; At the LHC, $4.4 \times 10^6$ $\Xi_{bc}$ and $1.3 \times 10^5$ $\Xi_{bb}$ events can be generated via $p$-Pb collision, $3.5 \times 10^6$ $\Xi_{bc}$ and $1.0 \times 10^5$ $\Xi_{bb}$ events can be generated via Pb-Pb collision. The number of $\Xi_{bc}$ events to be generated at the RHIC or LHC are smaller than that of the $\Xi_{cc}$ events under the same collision by about one order; and the number of $\Xi_{bb}$ events to be generated at the RHIC or LHC are smaller than that of the $\Xi_{cc}$ events under the same collision by about two order. Those results show that if more experimental data have been accumulated at the RHIC and LHC, one may also have the chance to study the properties of the other two doubly heavy baryons, $\Xi_{bc}$ and $\Xi_{bb}$.

\section{Summary}

We have studied the $\Xi_{cc}$ production via $p$-N and N-N collisions at the RHIC and LHC. The generator GENXICC with suitable changes for the use of nuclear PDF has been adopted for the calculation. Our results show that in addition to the gluon-gluon fusion mechanism, the extrinsic charm mechanisms via $g+c$ and $c+c$ subprocesses, are important to achieve a sound prediction for the $\Xi_{cc}$ production. By summing up contributions from $g+g$, $g+c$ and $c+c$ mechanisms and contributions from different spin-and-color configurations of the intermediate $(cc)$-diquark together, we observe that sizable number of $\Xi_{cc}$ events can be produced via $p$-N and N-N collisions at the RHIC and LHC. More explicitly, we have shown that $1.6 \times 10^6$ and $6.3 \times 10^6$ $\Xi_{cc}$ events can be produced via $p$-Au and Au-Au collisions at the RHIC, respectively; $8.1 \times 10^7$ and $6.7 \times 10^7$ $\Xi_{cc}$ events can be produced via $p$-Pb and Pb-Pu collisions at the LHC, respectively. Sizable number of $\Xi_{cc}$ events can be accumulated via $p$-N and N-N collisions at the RHIC and LHC. Thus, in addition to the $pp$ collision as now been performed by LHCb experiment, the $p$-N and N-N collisions at the hadron colliders can be a good platform for investigating the properties of the doubly heavy baryons. As shown by Figs.(\ref{ptdr}, \ref{ptdl}), the $g+c$ mechanism is larger than $g+g$ mechanism in small $p_t$ region via $p$-N and N-N collisions. If the experimental measurements can be extended to small $p_t$ region, then one may study extrinsic charm mechanism by carefully measuring the $\Xi_{cc}$ events.

\hspace{1cm}

\noindent{\bf Acknowledgements} This work was supported in part by the Natural Science Foundation of China under Grant No.11605029, No.11675239, No.11535002 and No.11625520, the science project of colleges and universities directly under the Guangzhou Education Bureau No.1201630158, the Foundation for Fostering the Scientific and Technical Innovation of Guangzhou University, and by the Fundamental Research Funds for the Central Universities under the Grant No.2018CDPTCG0001/3.

\end{document}